# Self-recording and manipulation of fast long-range hydrogen diffusion in quasi-free magnesium

Xiaoyang Duan,[1,2,†] Ronald Griessen,[3,†,*] Rinke J. Wijngaarden,[3] Simon Kamin,[1] and Na Liu[1,2,*]

[1]*Max Planck Institute for Intelligent Systems, Heisenbergstrasse 3, 70569 Stuttgart, Germany.*
[2]*Kirchhoff Institute for Physics, University of Heidelberg, Im Neuenheimer Feld 227, 69120 Heidelberg, Germany.*
[3]*Department of Physics and Astronomy, Faculty of Science, VU university, Amsterdam, De Boelelaan 1081, 1081 HV Amsterdam, The Netherlands.*

[†]These authors contributed equally to this work.
[*]Corresponding authors: r.p.griessen@vu.nl, laura.liu@is.mpg.de

Understanding diffusion of large solutes such as hydrogen and lithium in solids is of paramount importance for energy storage in metal hydrides and advanced batteries. Due to its high gravimetric and volumetric densities, magnesium is a material of great potential for solid-state hydrogen storage. However, the slow hydrogen diffusion kinetics and the deleterious blocking effect in magnesium have hampered its practical applications. Here, we demonstrate fast lateral hydrogen diffusion in quasi-free magnesium films without the blocking effect. Massive concomitant lattice expansion leads to the formation of remarkable self-organized finger patterns extending over tens of micrometres. Detailed visualization of diffusion fronts reveals that the fingers in these patterns follow locally the direction of hydrogen diffusion. Thus, the streamlines of the diffusion process are self-recorded by means of the finger pattern. By inclusion of fast hydrogen diffusion objects or local gaps, the resulting streamlines exhibit a clear analogy to optical rays in geometric optics. The possibility to spatially manipulate hydrogen diffusion opens a new avenue to build advanced hydrogen storage systems, cloaking and active plasmonic devices, as well as prototype systems for computational models.

Keywords: hydrogen diffusion, magnesium hydride, self-organized patterns, self-recording, energy storage, solute-induced swelling

## I. INTRODUCTION

The magnesium-hydrogen (Mg-H) system has constantly fascinated researchers due to its interesting hydrogen storage properties. With relatively high gravimetric and volumetric densities (7.6 wt% and 110 g l$^{-1}$, respectively), inexpensive and abundant Mg is one of the most attractive metals for solid-state hydrogen storage [1,2]. The relentless interest in Mg-H is, however, also due to a large extent to the high thermodynamic stability of $MgH_2$ as well as its slow and complex H absorption/desorption kinetics. Despite an impressive body of information [2–4], the published values for H diffusion in Mg exhibit a large spread varying from $10^{-20}$ up to $10^{-15}$ m$^2$s$^{-1}$ near room temperature [5,6]. Detailed studies of H absorption and desorption in Mg thin films [5–10] have shown that the sluggish kinetics and the resulting notorious 'blocking effect' are mainly due to the formation of a superficial layer of the extremely stable hydride [2], $MgH_2$. This has led to the belief that fast H diffusion in concentrated Mg-H systems is inherently impossible.

Here we demonstrate that the blocking effect is absent, when H enters a Mg film laterally, *i.e.*, parallel to the film plane as indicated in Fig. 1. We further show that in such a lateral diffusion geometry a quasi-free Mg film has the possibility to minimize the very large H-induced lattice dilation by a massive modification of its morphology. This generates self-organized finger patterns that extend over tens of micrometres along the diffusion direction. These finger patterns are optically monitored in real time and can be used for simultaneous measurements of H diffusion coefficients in a large number of samples. The corresponding finger landscapes shown in atomic force microscopy (AFM) images reveal the H streamlines during diffusion, which can be manipulated by spatial modulations using objects with different H diffusion coefficients. The existence of H diffusion finger patterns and the possibility to manipulate them have never been reported thus far. Especially interesting is that the diffusion process is self-recording its progress in Mg by means of the finger pattern.

## II. EXPERIMENT

### 1. Samples

Each sample consists of a Pd entrance gate or multiple Pd gates (typically 20 nm thick) and a layer of Mg with thickness varied between 22 and 372 nm. In all samples, a 3 nm Ti buffer layer is deposited on a $SiO_2$ (100 nm)/Si substrate to minimize the clamping of the Mg layer to the substrate [11–16]. An oxidized $TiO_x$ capping layer covers the Mg layer so that H absorption or desorption can only occur via the Pd gate(s). The Ti buffer layer, the Mg layer, and the Ti capping layer are successively deposited on the substrate through electron-beam evaporation (PFEIFFER Vacuum, PLS-500). The deposition rates of Mg and Ti are 1.0 and 0.2 nm/s, respectively. The vacuum and temperature during depositions are $1\times10^{-5}$ Pa and ~20 °C, respectively.

The samples are manufactured using multiple electron-beam lithography (EBL). First, alignment markers are defined in a double layer PMMA resist (200k-3.5% and 950k-1.5%, Allresist) using EBL (Raith e_line) on the substrate. A 2 nm chromium adhesion layer and a 40 nm Au film are deposited on the substrate using thermal evaporation followed by a standard lift-off procedure. Next, the substrate is coated with a double layer PMMA. Computer-controlled alignment using Au markers of the first layer are carried out to define a structural layer. Subsequently, 20 nm Pd for the hydrogen entrance gates and fast H-diffusion objects (*e.g.*, Pd prisms and lenses) is deposited on the substrate using thermal evaporation followed by the same lift-off procedure. Then, the substrate is coated with a double layer PMMA again, and the alignment Au markers are used to define another structural layer. After development (90 s in MIBK and 60 s in isopropanol), 3 nm Ti, 22–372 nm Mg, and 5 nm Ti are successively deposited on the substrate through electron-beam evaporation (PFEIFFER Vacuum, PLS-500).

## 2. Optical measurements

During hydrogen absorption or desorption optical images and videos of the samples are taken using a bright-field reflection microscope (Nikon, ECLIPSE LV100ND) illuminated by a white light source (Energetiq Laser-Driven Light Source, EQ-99). A digital CCD Camera (Allied-Vision Prosilica GT2450C) is used to capture images with a ×100 (NA = 0.6) objective. The hydrogenation and dehydrogenation experiments are carried out in a homemade optical gas chamber. The flow rate of the gas (hydrogen or oxygen) is 2.0 l min$^{-1}$. The temperature of the cell can be varied between room temperature and 100 °C.

## 3. AFM and SEM measurements

The topographies of all samples are measured with an AFM instrument (MultiMode 8) equipped with NanoScope V controller, vertical J-scanner, and NanoScope version 8 software (Bruker AXS, Santa Barbara, CA). AFM measurements are performed using Si cantilevers (NANOSENSORS, PointProbe-Plus) with scan size of 40 μm, scan rate of 0.5 Hz, and resolution of 1024 × 1024. The SEM images are taken using an ULTRA 55 system (Carl Zeiss AG) at 15 kV.

## 4. Focussed-ion beam measurements

Cross-sections of the samples are generated by Focused Ion Beam (FIB) system (Raith FEI Multibeam equipped with a gallium ion beam etching function) to check that the topographic finger patterns observed in AFM are not due to delamination or buckling. To observe a fresh surface, the sample is tilted 52° to perform ion milling with a gallium ion beam current of 10 pA operating at 30 kV, in a series of steps across the surface, at increasing depths, until the sample milled into the substrate.

## III. UBIQUITY AND GENERAL PROPERTIES OF FINGER PATTERNS

We have observed finger patterns in all our samples. Three representative examples are shown in Fig. 1 (see Appendix A for details). Regardless of whether H is introduced via a narrow underlying Pd strip [see Fig. 1(a)], a Pd ring at the periphery of a Mg disk [see Fig. 1(b)], or a Pd 'star' consisting of 6 radial strips [see Fig. 1(c)], in all cases finger patterns are observed both in AFM and optical reflection (OR) images. There is a remarkable one-to-one correspondence between the AFM and OR finger patterns. While a radial finger pattern is expected due to the symmetric configuration in Fig. 1(b), the patterns in Figs. 1(a) and 1(c) show that the fingers tend to nucleate in a direction perpendicular to the radial Pd strips. For different samples, the finger patterns exhibit universal characteristics, although the details vary in individual samples. With increasing distance from the Pd entrance gate, i) the average separation between two adjacent fingers increases; ii) the width of the individual fingers increases; iii) the sample height averaged along lines perpendicular to the H streamlines remains constant; iv) the out-of-plane expansion averaged over the entire finger pattern region is essentially equal to the total volume expansion (see Appendix B for more details).

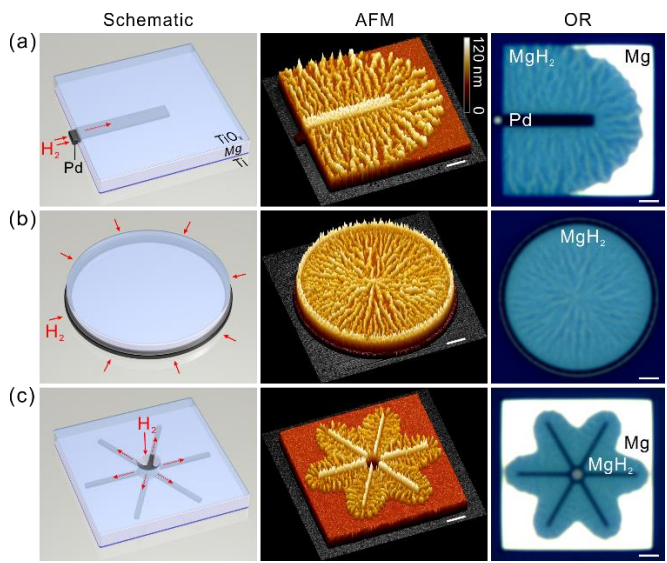

FIG. 1. H enters quasi-free Mg through (a) a 1 μm wide Pd strip under a 3 nm Ti/45 nm Mg/5 nm TiO$_x$ patch, (b) a 0.6 μm wide Pd ring at the periphery of a Mg disk, and (c) a 1.5 μm opening in a Mg patch with 6 radial Pd strips of 0.2 μm width underneath. Schematics of the samples are shown in the left column. The red arrows indicate how H enters the samples and then diffuses parallel to the Mg film plane. The AFM landscapes and the corresponding OR images taken after H loading of (a) 28,350 s, (b) 11,580 s, and (c) 1,390 s at 353 K at a hydrogen pressure of 20 kPa. The light blue regions in the OR images of (a), (b), and (c) correspond to MgH$_2$, while the bright regions in the OR images of (a) and (c) are due to metallic Mg. (Scale bar: 2 μm).

To unravel the nature of the finger patterns and unambiguously identify the involved hydride phases, we consider the simple geometry shown in Fig. 2(a). The 15 × 15 μm$^2$ sample consists of a 45 nm thick Mg film sandwiched between a 3 nm Ti buffer layer and a 5 nm TiO$_x$ capping layer.

H enters the sample via a 20 nm thick and 15 μm long Pd strip spanning the entire left-hand side of the Mg patch. The AFM, OR, and scanning electron microscopy (SEM) images are recorded after H loading at pressure of 20 kPa at 353 K. It reveals that the AFM, OR, and SEM finger patterns are strikingly similar.

At first sight one might naively associate the formed hills with MgH$_2$ and the plains with Mg, respectively. Reality is, however, quite different. As shown by the OR image in Fig. 2(a), the hills and plains are optically yellowish and bluish, respectively. Both are much darker than the region ahead of the finger pattern. This indicates that the entire finger pattern, *i.e.*, both hills and plains, occurs within the insulating MgH$_2$ phase and the region ahead of the pattern is essentially metallic Mg. The various shades in blue colour are due to interference effects in the transparent dihydride [17,18], whose thickness is strongly position dependent. As evidenced by the AFM image, the sample landscape comprises plains and hills (*i.e.*, the fingers). The hills can be approximately 200% higher than the plains. An essential clue for the interpretation of such high hills is provided by the SEM measurements in Fig. 2(a), which show within 0.5% accuracy there is no in-plane dilation upon hydrogenation. A direct consequence of this observation is that the out-of-plane dilation averaged over the entire finger pattern measured with AFM is expected to be approximately 30%, as the molar volumes of Mg and MgH$_2$ are 13.98 and 18.15 cm$^3$mol$^{-1}$, respectively. This expectation is confirmed by the measured average heights derived from the height histogram of the sample in Fig. 2(b).

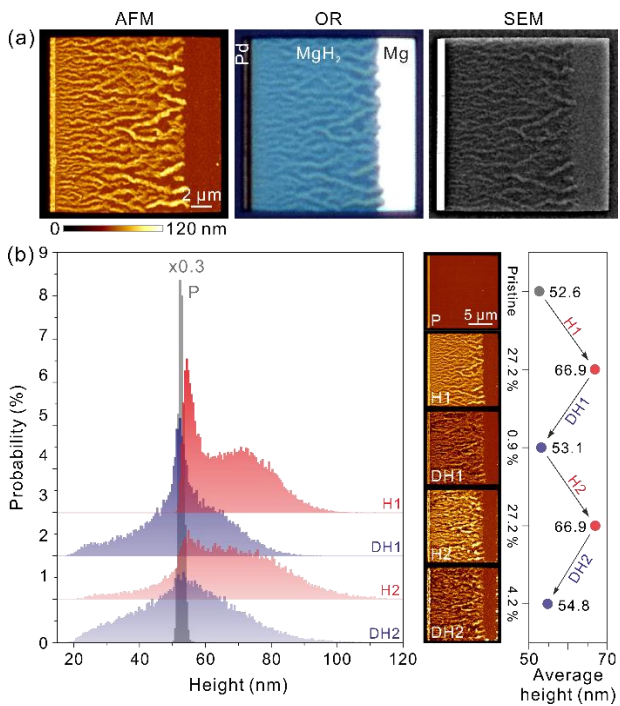

FIG. 2 (a) AFM, OR, and SEM images of the finger patterns after the first hydrogenation (41,390 s) at 20 kPa at 353 K in a 3 nm Ti/45 nm Mg/5 nm TiO$_x$ sample with a 20 nm thick Pd entrance strip. The light yellowish and bluish colours in the OR image correspond to the hills and plains in the AFM image, respectively. The hills and plains are all in the MgH$_2$ phase. The metallic Mg phase appears very bright in the OR image to the right side of the finger pattern. (b) Histograms of the finger patterns for the pristine (P), first hydrogenation (H1), first dehydrogenation (DH1), second hydrogenation (H2), and second dehydrogenation (DH2) states sequentially. The histograms are all calculated within the same region of the AFM images, which is determined by the extent of the finger pattern in the hydrogenated zone (approximately 15 × 11 μm$^2$). The histogram P is scaled by a factor 0.3 for clarity. For each state, the average height derived from the corresponding histogram is indicated in the right panel. These average heights are in excellent agreement with the predicted values from a model assuming that the total volume expansion is entirely accommodated out-of-plane.

For an unambiguous identification of the phases involved in the finger patterns, we have characterized this sample using AFM for two H loading/unloading cycles in Fig. 2(b). The sharply peaked histogram of the pristine state (P) in Fig. 2(b) indicates that the surface of the as-fabricated sample is flat with an average height of 52.6 nm. After H loading for 41,390 s at 20 kPa, the histogram of the first hydrogenation (H1) exhibits two peaks. The peak centred at ~54.2 nm arises from the plains of the landscape, while the broad peak at ~70.8 nm is due to the hills, whose maximum height can be larger than 100 nm. Most interesting is that the average height derived from histogram H1 is 66.9 nm. After the first unloading, the average height derived from histogram of the first dehydrogenation (DH1) is 53.1 nm. The two measured average heights are in excellent agreement with the values 67.17 nm and 53.75 nm predicted by assuming that the volume changes from Mg to MgH$_2$ (29.83%) and from Ti to TiH$_2$ (24.95%) are entirely accommodated out-of-plane (see part 2 of Appendix B for details). The average height 53.1 nm measured after the first unloading is also in excellent agreement with 53.75 nm since the much larger stability [19] of TiH$_2$ compared to MgH$_2$ prevents it from unloading. In the DH1 state, the sample therefore consists of a TiH$_2$ buffer layer and a metallic Mg film. Further justification of this interpretation is that upon the second H loading the average height reaches again 66.9 nm. After the second H unloading, the slightly higher average height 54.8 nm is probably due to an incomplete dehydrogenation. An additional proof of the vanishing in-plane expansion upon H loading follows directly from the SEM analysis of the samples containing various amounts of hydrogen as indicated in part 3 of Appendix B.

That the entire finger pattern consists of MgH$_2$ is further substantiated by two essential observations: i) after the first unloading, histogram DH1 is almost symmetric and broad, ranging from 20 to 80 nm. This implies, as shown in the AFM images of Fig. 2(b), that hills have been lowered. More important is that deep crevasses have developed between the hills. This is only possible if, after the first H loading, the plains are also fully hydrogenated; ii) the dark and light blue colours in the OR image of Fig. 2(a) are fully consistent with interference colours through a transparent layer of MgH$_2$.

For all samples investigated in this work, we have observed well-developed finger patterns exhibiting the same characteristic features. The finger patterns are generated in order to minimize the internal stresses in the Mg-H system that must expand by approximately 30% in volume, while keeping its in-plane dimensions constant. Very important is to realize that the constancy of in-plane dimensions cannot be due to a strong sticking of the Mg film to the substrate, as Mg does not alloy with the Ti buffer layer. It is thus an intrinsic property of the quasi-free thin Mg films [11–16], which are in resemblance to wrinkling of skins [20].

## IV. MANIPULATION OF DIFFUSION STREAMS

The three representative examples shown in Fig. 1 demonstrate clearly that the finger patterns depend on the geometries of the samples and the Pd entrance gates. This immediately suggests that H diffusion in composite samples with spatially modulated diffusion coefficients should lead to specific distortions of the finger pattern presented in Fig. 2(a). Typical examples of such composite samples are shown in Fig. 3. The leading idea of the sample architecture is to include an object with H diffusion properties that are markedly different from those of the surrounding Mg film. For example, in the schematic of Fig. 3(a), a Pd prism is embedded in the Mg film. At the temperatures of interest in this work, *i.e.*, between 20 °C and 100 °C, H diffuses through Pd several orders of magnitude faster than through Mg. Both the AFM and OR images in Fig. 3(b)-*i* confirm that the finger pattern is drastically modified due to the presence of the Pd prism. The same also evidently happens for the convex and concave lenses shown in Fig. 3(b)-*ii*,-*iii*. On the other hand, a control experiment with a gold prism embedded in Mg [see Fig. 3(c)] confirms that the finger pattern is unaffected by an object that does not absorb H.

The observed finger patterns in Fig. 3(b)-*i* to *iii* seem to be counterintuitive, as it is well known in geometric optics that light is focused by convex lenses and defocused by concave lenses. One should however realize that conventional optical lenses are made of materials with refractive indices larger than 1, *i.e.*, $n_{object} > n_{medium}$. In our case, however, the Pd objects locally enhance the H diffusion. Here the geometric optics analogue is $n_{object} < n_{medium}$. Schematic representations of the geometric optical properties of such objects are indicated in Fig. 3(d). Another type of objects is small gaps in Mg films, generated by locally removing materials throughout the entire sample thickness. The finger patterns for the single gap and double slits in Fig. 3(b)-*iv*,*v* resemble the patterns encountered in wave optics. The clear analogy between H diffusion streams and optical rays is a direct consequence of the observation that Snell's law is approximately obeyed in diffusive processes [21,22].

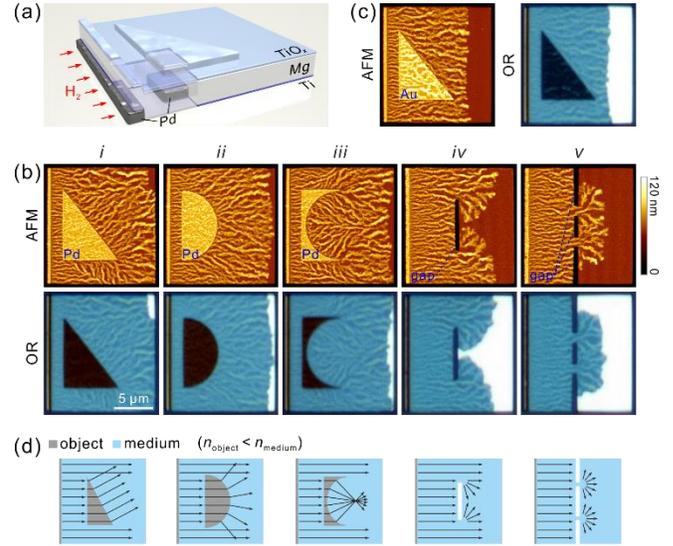

FIG. 3. (a) Schematic of a composite sample with a Pd prism under a 3 nm Ti/45 nm Mg/5 nm TiO$_x$ patch. A corner of the structure is made transparent for better visibility of the embedded 20 nm thick Pd prism. The red arrows indicate that H enters the sample via the Pd strip and diffuses parallel to the Mg film plane. (b) AFM and OR images of the Mg patches with various embedded Pd objects (characterized by a fast H diffusion) in *i*, *ii*, and *iii*, or with local air gaps (corresponding to impenetrable H diffusion obstacles) in *iv* and *v* (c) AFM and OR images of a control sample, in which an Au prism is embedded in a Mg patch. In this case, the finger pattern is not affected by the Au object, since Au does not absorb H. For all the samples, H loading is carried out under the same experimental conditions as in Fig. 1. (d) Analogy between lateral H diffusion in Mg and optical wave behaviour. Prism (*i*), convex (*ii*), and concave (*iii*) lens-like behaviour is obtained using geometric optics analogues with $n_{object} < n_{medium}$. H circumvents the single gap in *iv* and diffuses through the two slits defined by three gaps in *v*.

## V. SELF-RECORDING AND SIMULATIONS OF H STREAMLINES

So far, we have only presented the results obtained *after* H loading/unloading by means of AFM, OR and SEM characterizations. As a matter of fact, we have also recorded the OR videos of the diffusion processes *in real time* (see Appendix C). As the MgH$_2$ phase is much darker than the metallic Mg phase in OR, the recorded videos contain detailed information about the time dependence of the H diffusion front through a sample. This has led to the striking discovery that the diffusion fronts extracted from snapshots of the OR video at different times are locally orthogonal to the fingers. It is explicitly demonstrated in Fig. 4(a) for the five cases in Fig. 3(b). The fact that the sample unveils its H streamlines by itself in the form of a finger pattern due to a self-organized corrugation of its surface amounts to a self-recording of the entire diffusion process. In other words, the fingers are always generated along a direction locally perpendicular to the momentary diffusion front.

All the salient features of the finger patterns in Fig. 3(b) can be nicely reproduced by the diffusion model described in detail in Appendix E. They are based on numerical solutions to the diffusion equation

$$\frac{\partial c}{\partial t} = D_i \nabla f(c) \cdot \nabla c + D_i f(c) \nabla^2 c, \qquad (1)$$

where $t$ is the time and $D_i$ is the amplitude of the H concentration dependent diffusion coefficient. The index $i$ stands for medium (Mg) or object (Pd underneath Mg). The H concentration $c$ dependent function $f(c)$ is chosen for all simulations as

$$f(c) = e^{-\alpha c} + e^{-\alpha(1-c)}, \qquad (2)$$

with $\alpha = 20$ derived using a lattice-gas model for H in a metal. The symmetry of $f(c)$ is a direct consequence of the observed H loading/unloading symmetry (see Appendix C), which proves unambiguously that all the effects observed in this work result from genuine H diffusion and not from uncontrolled spurious effects. Important is to stress that the simulations involve only one adjustable parameter, $D_{object}/D_{medium}$, which is the same for all simulations shown in Fig. 4(c). An excellent agreement between the finger patterns in the AFM images and the simulated streamlines [black lines in Fig. 4(c)] is obtained with $D_{object}/D_{medium} = 100$. At time $t_0$, the H concentration in Mg is set equal to $c_0$ at the Pd entrance gate, while it is zero everywhere else. From this initial situation, Eq. (1) is numerically solved. At times $t_n \propto n^2$, where $n$ is an integer, the diffusion front lines (red lines) are plotted in Fig. 4(c). Meanwhile, the streamlines that are locally orthogonal to the front lines are constructed. The good agreement between experimental finger patterns and simulated streamlines confirm that the fingers are in fact a self-recording of the entire diffusion history.

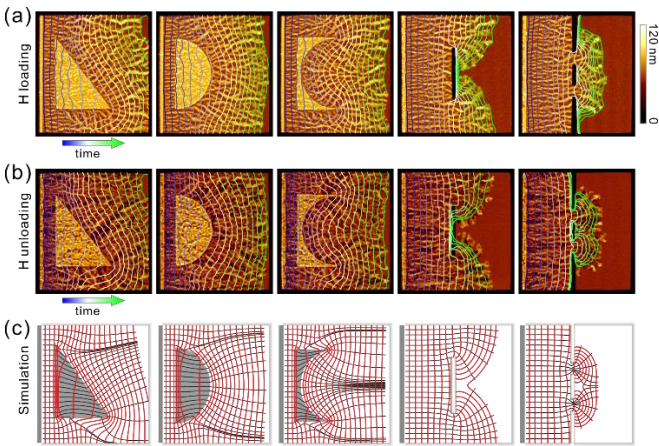

FIG. 4. (a) Diffusion fronts (blue to green lines) extracted from OR videos at times $t_n$ so that $t_n^{1/2}$ are equally spaced. The diffusion front lines are superimposed on the AFM images of the samples in Fig. 3(b), taken after H loading. It demonstrates that the fingers are locally orthogonal to the diffusion fronts. (b) Diffusion fronts (blue to green lines) extracted from OR videos during H unloading are superimposed on the AFM images of samples in Fig. 3(b), taken after H unloading. The similarity between the recorded diffusion fronts during H loading (a) and unloading (b) demonstrates the reversibility of the diffusion processes and implies the symmetric form of the function in Eq. (2). (c) Diffusion fronts (red lines) and streamlines (black lines) obtained from the simulations described in Appendix E are in excellent agreement with the experimental data in (a) and (b).

## VI. SYNERGETIC INTERACTION OF FINGER PATTERNS

An interesting feature of the time dependence of the measured diffusion front positions (see part 1 of Appendix D for details) is that the square root of time behaviour occurs only after a time delay of approximately 140 s at 353 K. This is the characteristic signature of a diffusive process [23] following a nucleation step. A close inspection of the AFM images in the immediate vicinity of the Pd entrance strip reveals a full row of regularly spaced hills, somewhat reminiscent of the patterns observed in compressed thin films on compliant substrates [24]. The observed time delay is due to the nucleation of these primordial hills. Once formed, they serve as seeds for the finger pattern generated by the continuous H absorption. While the H diffusion front moves through the sample, there is a competition between fingers. This leads to a slow increase of the width of the dominant fingers, while other fingers are terminated. The evolution of the finger pattern has been observed in all samples investigated in this work. It is obviously energetically more favourable to extend existing hills rather than generating new ones. Hydrogen is a dilation center in metals (see Appendix B). Therefore, at the tip of a finger, where the lattice is dilated out-of-plane, it is more favourable for H to nucleate and form a new dihydride in this region. This is at the origin of the self-organisation of the finger patterns. A vivid demonstration of the favoured hill nucleation at the tip of fingers is shown in Fig. 5. By using several Pd entrance gates for H diffusion, finger patterns can collide with each other. For example in Fig. 5(a)-*iii*, finger patterns generated from two opposite directions collide at the median line. The head-on collision of two fingers leads to the synergetic generation of a huge central ridge. This synergetic effect, although weakened when two fingers meet under a sizeable angle (for example 90°), leads to the formation of meeting-lines that depend on the sample geometry and the geometric arrangement of the Pd entrance gates. This is clearly seen in the AFM images from Fig. 5(a)-*ii* to *v*. Although the finger patterns are largely influenced by the positions and the number of the Pd entrance gates, the height increase (averaged over the entire patch) generated by lateral H absorption is remarkably constant and close to 27.2%, which is the value measured in Fig. 2. The ridges highlighted in the dashed boxes in Fig. 5(a) can reach relative height changes as high as 300%, which slightly increases the average height values from *i* to *v* stepwisely. The experimental finger patterns agree very well with the simulated streamlines (black lines) as shown in Fig. 5(b).

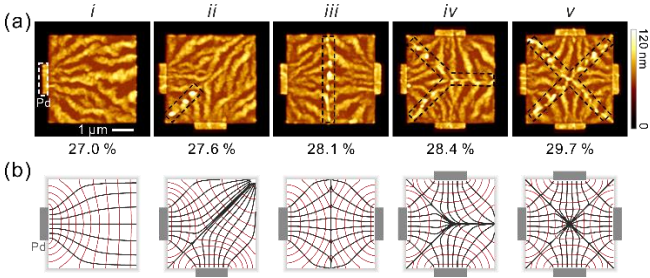

FIG. 5. (a) AFM images of the finger patterns for different arrangements of the Pd entrance gates. The samples are all $4 \times 4$ μm² in size. The thicknesses of the individual Mg and Ti layers are the same as those in the previous figures. The black-dashed boxes highlight the high ridges resulting from the synergetically enhanced local piling up of Mg. The relative height changes averaged over the entire individual samples are (*i*) for one Pd gate, 27.0%; (*ii*) for two perpendicular Pd gates, 27.6%; (*iii*) for two parallel Pd gates, 28.1%; (*iv*) for three Pd gates, 28.4%; and (*v*) for four Pd gates, 29.7%. The ridges highlighted in the dashed boxes can reach relative height changes as high as 300%, which increases the average height values from *ii* to *v* stepwise. (b) Simulation results for the sample configurations in (a). The diffusion fronts (red lines) are obtained from the simulations described in Appendix E. They are obtained for a chosen set of times $t_n$ so that $t_n^{1/2}$ are equally spaced. The streamlines (black lines) are locally orthogonal to the diffusion fronts (red lines).

## VII. HYDROGEN DIFFUSION MEASUREMENTS

The recorded OR videos used to demonstrate the H loading/unloading symmetry in Fig. 4 are also directly relevant to measurements of the front mobility itself. The sharp colour change at the front of the finger patterns makes it possible to monitor accurately the time dependence of the diffusion front position. The results in part 1 of Appendix D show that the square of the front position $x_{\text{front}}^2$ is proportional to time $t$, after the original finger seeds have been nucleated at $t_{\text{nucleation}}$. The front mobility $K$ is defined as

$$K = \frac{x_{\text{front}}^2}{t - t_{\text{nucleation}}}, \quad (3)$$

and has the same dimension (m²s⁻¹) as a diffusion coefficient [23]. The great advantage of the optical hydrogenography is that it allows to monitor simultaneously a large number of samples [25], therefore reducing statistical errors to an unprecedented minimum. The results of the samples with various Mg thicknesses in Fig. 6 exhibit a series of absolutely remarkable features. First, in-plane H diffusion is fast. It is almost five orders of magnitude faster at 300 K than the value of out-of-plane diffusion from Spatz *et al*. [5]. It remains fast over macroscopic distances in sharp contrast to the results of Uchida *et al*. [6], who observed fast diffusion only during a relatively short period of time (~1,400 s) over a short distance (~200 nm) before the blocking layer was formed. In their case, the formation of a 200 nm thick blocking layer reduced further H absorption by approximately two orders of magnitude and the corresponding apparent diffusion coefficient dropped to typically $10^{-18}$ m²s⁻¹. Second, the temperature $T$ dependence of the front mobility follows accurately the Arrhenius relation

$$K = K_0 \exp\left(-\frac{E_a}{k_B T}\right), \quad (4)$$

with the prefactor $K_0 = 1.11 \times 10^{-9}$ m²s⁻¹, and the activation energy $E_a = 0.389$ eV independent of the Mg thickness. $k_B$ is the Boltzmann constant. The value of $E_a$ is close to that of H in Mg alloyed [26] with 2 at% cerium, where $E_a = 0.41$ eV. It is also comparable to that for H diffusion in the fcc metals Cu (0.40 eV), Ni (0.40 eV), Pd (0.23 eV) [27] and the hcp Y (0.37-0.38 eV) [28]. Comparison with hcp and fcc metals is meaningful as these two structures differ only by the stacking of closed packed hexagonal planes.

Third, in contrast to perpendicular diffusion experiments where relatively low $H_2$ pressures must be used in order to minimize the deleterious effects of the $MgH_2$ blocking layer [6,8], we find for lateral H diffusion that the front mobility increases with increasing hydrogen pressure as indicated in part 2 of Appendix D.

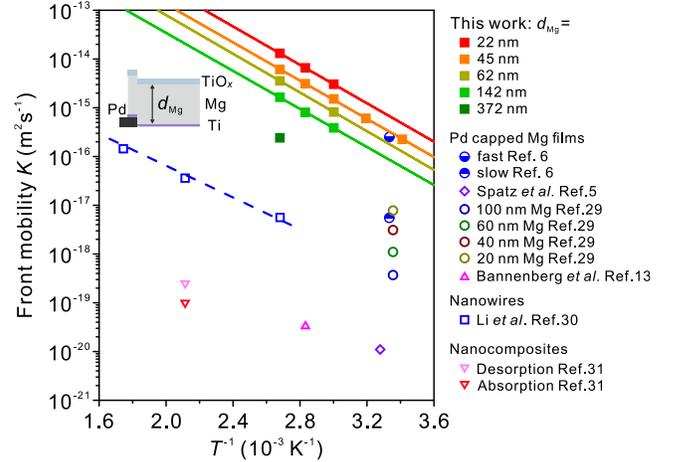

FIG. 6. Lateral front mobilities measured optically on 18 equivalent 3 nm Ti/$d_{Mg}$ Mg/5 nm TiO$_x$ samples with $d_{Mg}$ varying from 22 to 372 nm (filled squares). For all the samples, the Pd entrance gate is 20 nm thick (see the inset schematic). The activation energy determined from the slope of the Arrhenius plot for the sample with $d_{Mg} = 45$ nm is $0.389 \pm 0.005$ eV. Considering the fact that the front mobility depends only on the ratio of the Mg and Pd entrance strip thicknesses (see Fig. 14), the slopes of all Arrhenius plots should be the same. From the four Arrhenius plots in Fig.6, one finds an activation energy of 0.390 eV ± 0.003 eV. Estimated values of the front mobilities for Pd capped Mg films [5,6,13,29], nanowires [30], and nanocomposites [31] are also included. The values for H diffusion in a very thin Mg layer sandwiched between relatively thick Ti and Pd layers [13], in 20~100 nm Mg films capped with a 5 nm Pd layer [29], in nanowires [30], and in nanocomposites [31] are estimated from the published absorption or desorption curves using $K \propto L^2/t_{1/2}$, where $L$ is the characteristic size of the samples and $t_{1/2}$ is the time necessary to half-fill the samples. Interestingly, the slope of the Arrhenius plot for the nanowire data [30] is comparable to the slopes of our plots.

## VIII. CONCLUSIONS

Using a sample architecture where H diffuses laterally in a Mg film via Pd entrance gates, we have discovered that self-organized finger patterns made of $MgH_2$ can be generated over tens of micrometres. Their front mobility is large at all times and not hampered by the well-known 'blocking effect' generally observed in the published work [5–10], where H diffuses perpendicularly to Mg films. We have also demonstrated that the finger patterns are effectively a self-recording of the entire H diffusion history. The fast lateral H diffusion and the possibility to manipulate H streamlines by inclusion of different diffusion objects or local gaps enable entirely new possibilities for applications in hydrogen storage systems [1,2], active plasmonics [15,32–34] and cloaking [35,36]. Furthermore, as lithium is also a large dilation solute in electrodes, our work is relevant to electrical energy storage in batteries [37–41]. In particular, the very detailed information gathered in this work about the massive influence of H diffusion on the structure and morphology of Mg films can be advantageously used to evaluate the applicability of (chemo)mechanical multiscale computational models for describing lithiation processes in high-volume-change electrode materials [42]. Finally, it is noteworthy that there is a remarkable similarity between the H-induced finger patterns in Mg and the vortex avalanches in superconductors [43].

## ACKNOWLEDGEMENTS

We gratefully acknowledge the generous support by the Max Planck Institute for Solid State Research for the usage of clean room facilities. We also thank the 4th Physics Institute at the University of Stuttgart for using their electron-beam evaporation system. We thank Ulrike Eigenthaler and Michael Hirscher for helping with the focused ion beam system, as well as Wiebke Lohstroh for the initial optical simulations. This project was supported by the Sofja Kovalevskaja grant from the Alexander von Humboldt-Foundation, the Marie Curie CIG grant, and the European Research Council (ERC *Dynamic Nano*) grant.

## AUTHORS CONTRIBUTION

X.D., R.G. and N.L. conceived the study. X.D. designed and carried out the experiments, and analyzed the data. R.G. developed the interpretation framework and wrote the essential parts of the article. R.W. and R.G. developed the diffusion simulations. S.K. helped with the Mg evaporation. All authors discussed the results and commented on the manuscript.

## APPENDIX A: DETAILED SAMPLE LAYOUT

Here, we illustrate all the relevant dimensions of the samples in Fig. 1.

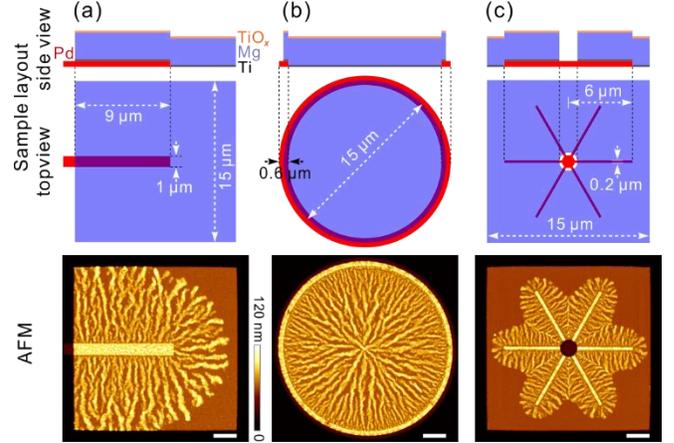

FIG. 7. Details of the layout of the three samples in Fig. 1. The AFM images are the same as in Fig. 1 but now in top view for a direct correlation with the sample layout. The 45 nm thick Mg layer is sandwiched between a 3 nm Ti buffer layer and a 5 nm $TiO_x$ capping layer. (Scale bar: 2 μm).

## APPENDIX B: REPRODUCIBILITY AND GENERAL PROPERTIES OF FINGER PATTERNS

### 1. Constancy of average height increase

Here, we indicate how the average heights $\bar{z}(x)$ calculated along the $y$-direction of the four samples in Fig. 8(a) depends on position $x$ with respect to the Pd-entrance gate. It is remarkable that $\bar{z}(x)$ is essentially independent of $x$ and the total average height increase $\overline{z(x)}$ is nearly the same for all four samples within 0.7%.

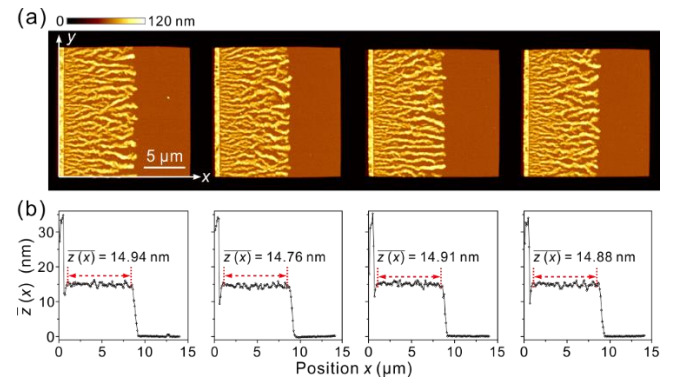

FIG. 8. (a) Finger patterns measured at 353 K after H loading at 20 kPa of $H_2$ in four identical $15 \times 15$ μm² 3 nm Ti/45 nm Mg/5 nm $TiO_x$ patches. The thickness of the Pd entrance gates are 20 nm. (b) Average heights calculated along the $y$-direction as a function of the $x$-position for the corresponding samples in (a). The total average

height increase $\overline{\overline{z(x)}}$ is the mean value of all line-averages $\overline{z}(x)$ evaluated within the finger pattern region in each sample.

### 2. Average heights for two loading/unloading cycles

The variations of the average sample height during the two loading/unloading cycles in Fig. 2 can fully be rationalized if one assumes that the bulk volume expansion of Mg and Ti upon H absorption is entirely accommodated out-of-plane. The molar volumes $V$ of Mg and MgH$_2$ are 13.98 and 18.15 cm$^3$mol$^{-1}$, respectively. The relative volume expansion of Mg is

$$\left(\frac{\delta V}{V}\right)_{Mg} = \frac{V_{MgH_2} - V_{Mg}}{V_{Mg}} = 29.83\% . \quad (B1)$$

For titanium, $V_{Ti} = 10.62$ cm$^3$mol$^{-1}$, $V_{TiH_2} = 13.27$ cm$^3$mol$^{-1}$ and

$$\left(\frac{\delta V}{V}\right)_{Ti} = \frac{V_{TiH_2} - V_{Ti}}{V_{Ti}} = 24.95\% . \quad (B2)$$

For the sample in Fig. 2, the pristine total thickness of design is 53 nm. After first H loading, the 3 nm Ti buffer layer and 45 nm Mg layer have formed dihydrides, while the 5 nm TiO$_x$ capping layer does not react with H. The resulting thickness predicted by Eqs. (B1) and (B2) is [3 nm × (1+24.95%) + 45 nm × (1+29.83%) + 5 nm] = 67.17 nm. This agrees nicely with the measured value 66.9 nm indicated for the first H loading in the right panel of Fig. 2(b). After the first H unloading, the large negative enthalpy of hydride formation of TiH$_2$ prevents the buffer layer from unloading and the predicted height is [3 nm × (1+24.95%) + 45 nm + 5 nm] = 53.75 nm, again in good agreement with the measured value of 53.1 nm. According to this model, the average heights of fully loaded samples should remain the same, *i.e.*, 66.9 nm, after the second loading and in fact for all subsequent loadings. This is nicely confirmed by the measured average height of 66.9 nm. Similar behaviour is expected for the unloaded samples. Fully unloaded samples should all have an average height of 53.75 nm. After the second unloading, the measured average height is however 54.8 nm, which is ~1 nm higher than after the first unloading. This could be due to the formation of voids. Cross-sections of the samples by means of the FIB did, however, not reveal blisters or buckling (see part 4 of Appendix B). More likely is that the sample was not fully unloaded after the second unloading. Nonetheless, it is absolutely remarkable that the assumption of a complete out-of-plane expansion during hydrogenation leads to average height estimates that are in very good agreement with the experiments.

### 3. Zero in-plane expansion during hydrogenation

The excellent agreement between the measured and predicted average heights in part 2 of Appendix B is *a posteriori* confirmation of the validity of the assumption that the volume change upon hydrogenation is entirely accommodated out-of-plane for the Ti/Mg/TiO$_x$ samples considered in this work.

Another proof of the vanishing in-plane expansion, can be obtained by looking simultaneously at patches with various overall hydrogen contents. The 9 patches in Fig. 9 have all the same dimensions. The only difference is the lengths of the Pd entrance gates that vary from 0.3 μm up to 5 μm. The SEM images are recorded at a time when the patch with the full length Pd gate (top right patch) is almost completely loaded with hydrogen, while only little hydrogen has entered the patch with the smallest Pd entrance gate (bottom left patch). The identical sample sizes (yellow and blue arrows) demonstrate that the lateral dimensions do not depend on the total amount of hydrogen. The total hydrogen content of a patch thus has obviously very little influence on the in-plane patch dimensions. Within experimental errors we conclude that for all the samples investigated in this work the lateral relative expansion is smaller than 0.5%.

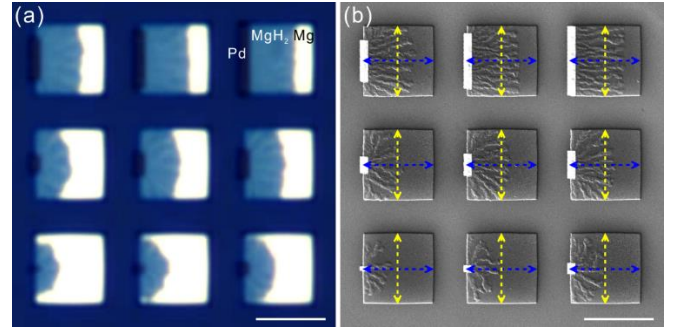

FIG. 9. (a) OR and (b) SEM images of 9 identical 3 nm Ti/45 nm Mg/5 nm TiO$_x$ samples with various lengths of the Pd entrance gates. The 5 × 5 μm$^2$ samples are simultaneously loaded with H for 1,390 s at 353 K at 20 kPa. The yellow and blue arrows in (b) are a guide to the eye to check that the lateral dimensions do not depend on the total amount of H. (Scale bar: 5 μm).

### 4. Absence of buckles

Cross-sections of the samples generated by Focused Ion Beam (FIB) show that the topographic finger patterns observed in AFM are not due to delamination or buckling. A representative example is given in Fig. 10.

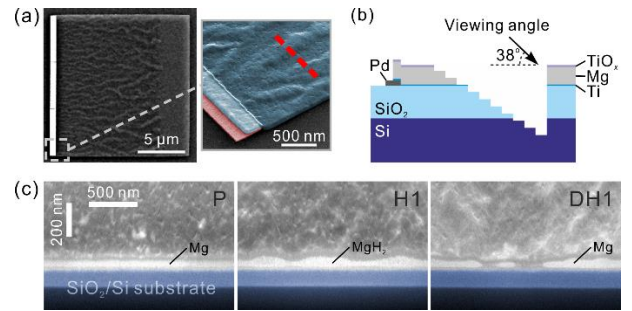

FIG. 10. (a) SEM images (top and detail of tilted view) of a representative sample. The red dash line indicates the position of the FIB milling. (b) Schematic of the FIB milling. (c) SEM cross-

sectional images of the samples for the pristine (P), first hydrogenation (H1), and dehydrogenation (DH1) states, respectively, after stepped ion-milling to expose the surface. The images are taken with a 38º tilt from horizontal to provide perspective on the vertically etched film surface.

## 5. Mechanism

The large lattice expansion induced by H absorption in metals can lead to delamination and buckling of metallic films evaporated on low adhesion substrates. This is for example the case for Niobium films evaporated on mica [44] or Pd films deposited on quartz [45]. The typical size of buckles is typically tens of micrometers. In our samples cross-sections obtained by means of FIB do not reveal any delamination or buckle. Furthermore, the width of individual fingers is always submicron.

The mechanism responsible for the accommodation of the large volume expansion (approx. 32%) accompanying H absorption in Mg is therefore of a different nature. It is related to the texture of our samples. Our Mg films have predominantly their hexagonal $c$-axis in a direction perpendicular to the substrate. During hydrogenation, rutile $MgH_{2-d}$ grows predominantly with a (110) direction perpendicular to the substrate. This is schematically indicated in Fig. 11. The resulting structure can be visualized as a structure made of quasi-hexagonal planes separated by $d(MgH_2) = 3.197$ Å with H occupying in-plane and inter-plane positions. The separation between hexagonal planes in the hcp structure of Mg being 2.065 Å we see that the out-of-plane expansion $[d(MgH_2) - d(Mg)]/d(Mg)$ upon H absorption in Mg is ~22.7%. The relative in-plane increase of the (quasi)hexagon area is only 8.3%. As a consequence stresses are efficiently minimized by nucleating the dihydride phase with a (110) direction perpendicular to the substrate. Delamination would destroy the Mg/Ti interface and cost additional surface energy [12]. In a lateral diffusion experiment the Mg layer has further the possibility to continuously minimize strains near the front. The overall result is that in our samples the lateral relative expansion is at most 0.5%. This is reminiscent of the morphological changes induced by H in another hexagonal host metal, Yttrium [46–48].

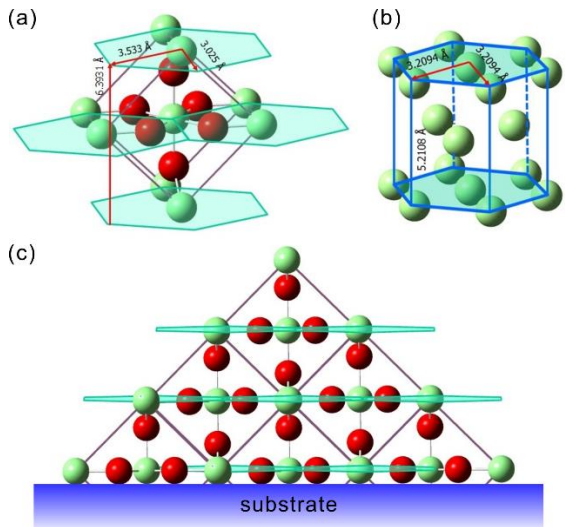

FIG. 11. (a) $MgH_2$ rutile unit cell with [110] direction perpendicular to the substrate. The quasi-hexagons are indicated in semi-transparent green. (b) HCP structure of Mg. (c) Part of the $MgH_2$ lattice viewed along the [001] direction of the rutile cell showing that H occupies both in-plane and inter-plane position of the Mg sublattice. The red and green balls represent H and Mg atoms, respectively.

## APPENDIX C: H LOADING-UNLOADING SYMMETRY

Snapshots taken from the videos of the H loading and unloading of the five samples in Fig. 3(b) are shown in Fig. 12. They illustrate nicely that the diffusion fronts during H absorption have the same shape as during H desorption at similar times after starting the absorption and starting the desorption, respectively. The times are not identical because different processes take place at the Pd entrance gate. During absorption, $H_2$ is catalytically split into two H atoms at the Pd surface, while during desorption two separate H atoms have to recombine to leave the Pd surface. The snapshots demonstrate also unambiguously that all the exchange of hydrogen with the Mg patches occurs via the Pd gates. During desorption, this is the reason why a bright reflection due to metallic Mg formation moves from left to right starting at the Pd gate.

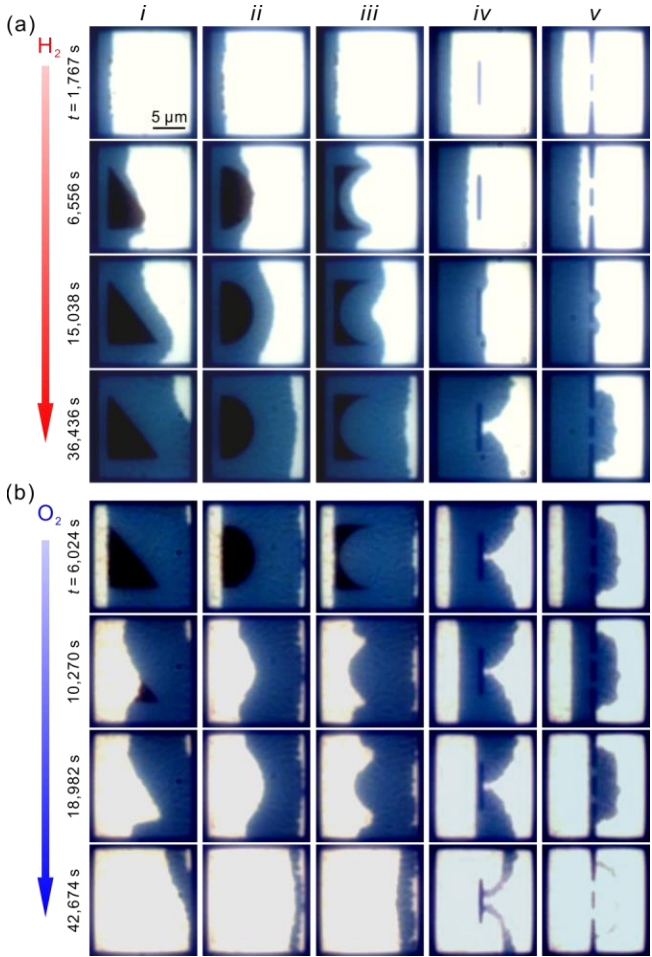

FIG. 12. (a) Snapshots of the OR video recorded at 353 K during H loading at 20 kPa of $H_2$ at times 1,767 s, 6,556 s, 15,038 s, and 36,436 s for the five samples in Fig. 3(b). (b) Snapshots of the OR video during H unloading at 353 K at 20 kPa of $O_2$ for the same samples taken at times 6,024 s, 10,270 s, 18,982 s, and 42,674 s, respectively. In all snapshot images, the dark bluish regions correspond to $MgH_2$, while the bright reflection areas are due to metallic Mg.

## APPENDIX D: MEASUREMENTS OF DIFFUSION FRONT MOBILITY

### 1. Temperature dependence of front mobility

Here, we indicate a typical example of parallel measurements. The sharp colour change at the front of the finger patterns makes it possible to monitor accurately the time dependence of the diffusion front position as shown in Fig. 13. The diffusion front positions separating the dark bluish $MgH_2$ finger regions from the bright metallic Mg are recorded with a camera and plotted as a function of time. The linear dependence of the squared front position as a function of time indicates the diffusive origin of the H motion. The slope of each line in the bottom panel is equal to the H diffusion front mobility $K$. The corresponding standard deviations are 3.51%, 3.51%, 5.25%, 4.93%, and 6.27%, respectively. Such an accuracy has never been demonstrated for H diffusion experiments between room temperature and 373 K. As a result the errors in the data points in Fig. 6 are much smaller than the size of the square symbols.

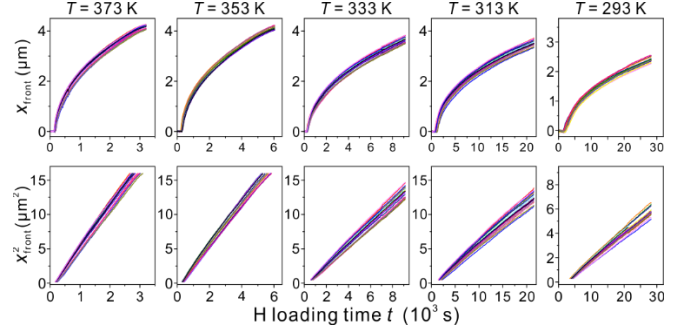

FIG. 13. Time dependence of the front position during H diffusion in 18 identical $5 \times 5$ μm$^2$ 3 nm Ti/45 nm Mg/5 nm TiO$_x$ patches loaded at 20 kPa of $H_2$ at different temperatures.

### 2. Dependence of the front mobility on the Mg thickness and $H_2$ pressure

Here we find that the front mobility depends on the thickness of the sample as shown in Fig. 14(a). For the sample architecture indicated in the inset of Fig. 14(a), one expects that the amount of hydrogen entering the 3 nm Ti/$d_{Mg}$ Mg/5 nm TiO$_x$ patch depends on the ratio of the thickness of the Pd entrance gate $d_{Pd}$ and that of the Mg $d_{Mg}$. To lowest order we expect that

$$K(d_{Pd}, d_{Mg}) \propto \frac{d_{Pd}}{d_{Mg}}. \tag{D1}$$

This expectation is confirmed by measurements of the front mobility for a series of samples with $d_{Pd} = 20$ nm and 22 nm $\leq d_{Mg} \leq 372$ nm. The front mobility varies linearly as a function of the inverse Mg thickness. Extrapolation of the data to $d_{Mg}^{-1} = 0.05$ leads to

$$K(d_{Pd} = d_{Mg} = 20 \text{ nm}) = 1.41 \times 10^{-14} \text{ m}^2\text{s}^{-1}. \tag{D2}$$

This is the front mobility of a sample with the same thickness of the Pd entrance gate and the Mg layer.

We also find that in contrast to perpendicular diffusion experiments where relatively low $H_2$ pressures must be used in order to minimize the deleterious effects of the $MgH_2$ blocking layer, for lateral H diffusion the front mobility increases with increasing hydrogen pressure as indicated in Fig. 14(b).

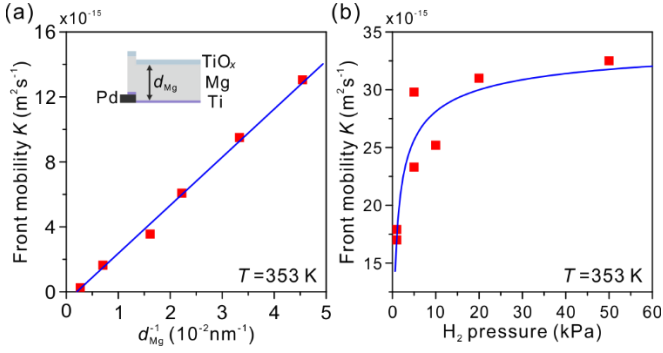

FIG. 14. (a) H diffusion front mobility measured on 3 nm Ti/$d_{Mg}$ Mg/5 nm TiO$_x$ samples at 353 K at 20 kPa. The thickness of the Pd entrance gate is always 20 nm while the thickness $d_{Mg}$ of the Mg layer varies from 22 to 372 nm. The observed linear dependence (blue line) confirms Eq. (D1). (b) H diffusion front mobility measured on 3 nm Ti/45 nm Mg/5 nm TiO$_x$ samples at 353 K at H$_2$ pressure varying from 1 to 50 kPa. The blue curve is obtained from a model based on the simulations described in Appendix E.

## APPENDIX E: THEORETICAL MODEL

The purpose of the simulations is to reproduce all the characteristic features of the finger patterns investigated in this work. To keep the simulations simple, the 3 nm Ti/$d_{Mg}$ Mg /5 nm TiO$_x$ samples are treated as one vertically homogeneous medium characterized with a hydrogen concentration dependent diffusion coefficient $D(c)$. We do not take into account the different crystal structures of Mg (hexagonal) and MgH$_2$ (rutile) and treat the Mg-H system as a lattice-gas MH$_c$ with $0 \leq c \leq 1$ and a critical temperature much higher than room temperature since the pressure-composition isotherms of Mg-H exhibit clear plateaus at all temperatures investigated here. Within this model the inclusion of objects with fast H diffusion (made for example of Pd) in the Mg layer is taken into account as a local modification of $D(c)$ that depends on the geometry of the inserted object. Excellent agreement with the observed finger patterns is obtained with a local enhancement factor of 100 at the object location for $D_i$ in Eq. (1).

### 1. Concentration dependent diffusion coefficient

The symmetry between H loading and unloading illustrated in Appendix C implies that the diffusion coefficient assumes similar values in the limit of small and high hydrogen concentrations. The concentration dependence chosen for all simulations is

$$D(c) = D_0 \left[ e^{-\alpha c} + e^{-\alpha(1-c)} \right], \tag{E1}$$

with $\alpha = 20$, where $D_0$ is a prefactor. Below we show that this relation, which is inspired by a simple lattice-gas model, leads to diffusion concentration profiles with a sharp front that mimics the coexistence of the dilute α-MH$_c$ (with typically $0 < c < 0.1$) and concentrated β-MH$_c$ (with typically $0.9 < c < 1$).

In the simplest mean-field description of a lattice gas the chemical potential of H in a metal M is

$$\mu_H = RT \ln\left(\frac{c}{1-c}\right) + H_H^\infty - Ac - TS_H, \tag{E2}$$

where $R$ is the gas constant; $T$ is the absolute temperature; $c$ is the H concentration, $c = $ H/M; $H_H^\infty$ is the enthalpy of H solution in M; $A$ is the magnitude of the long-range effective H-H interaction; and $S_H$ is the entropy of H in M. The minus sign in front of $A$, which is assumed to be a positive quantity (i.e., $A > 0$), corresponds to an attractive H-H interaction. At thermodynamic equilibrium with surrounding H$_2$ gas we have

$$\mu_H = \frac{1}{2}\mu_{H_2}. \tag{E3}$$

The chemical potential of H$_2$ gas is

$$\mu_{H_2} = H_{H_2} - TS_{H_2} = H_{H_2} - TS_{H_2}^0 + RT \ln p, \tag{E4}$$

where $p$ is the H$_2$ pressure expressed in unit of bar. Equations (E2) and (E4) together with (E3) lead to the pressure-composition isotherms [49]

$$\ln p = 2 \ln\left(\frac{c}{1-c}\right) + 2\frac{\Delta H_\infty - Ac - T\Delta S}{RT}, \tag{E5}$$

where

$$\Delta H_\infty = H_H^\infty - \frac{1}{2}H_{H_2}, \tag{E6}$$

and

$$\Delta S = S_H - \frac{1}{2}S_{H_2}^0, \tag{E7}$$

are the enthalpy and entropy of H solution in M, respectively.

The isotherm for the critical temperature $T_c$, has an inflexion point with zero slope, i.e., the first and second derivatives of $\ln p$ as a function of $c$ are both zero at the critical concentration $c = c_c$. From these conditions, we find

$$c_c = \frac{1}{2}, \tag{E8}$$

and

$$T_c = \frac{A}{4R}. \tag{E9}$$

The spinodal is by definition the curve, $T_s = T_s(c_s)$ for which

$$\frac{d}{dc}(\ln p) = 0. \tag{E10}$$

From Eq. (E5) with (E9) follows that

$$T_s = 4T_c c_s (1 - c_s). \tag{E11}$$

This simple relation shows that the spinodal curve is a parabola with its top at $c_c = 1/2$. Thus the spinodal is symmetric under the transformation $c = \delta \rightarrow c = 1 - \delta$ and at each temperature $T < T_c$ there are two spinodal concentrations $c_s^{low}$ and $c_s^{high}$: at temperatures below $T_c$ the dilute α-MH$_c$ and concentrated β-MH$_c$ coexist. The concentration $c_i$ of the coexisting α- and β-phases are given by the condition

$$\ln\left(\frac{c_i}{1-c_i}\right) - 4\frac{T_c}{T}\left(\frac{1}{2} - c_i\right) = 0, \quad \text{(E12)}$$

with $i$ = α-phase or β-phase. The coexistence curve $T_{\alpha\text{-}\beta} = T_{\alpha\text{-}\beta}(c)$ is then

$$T_{\alpha\text{-}\beta} = 4T_c \frac{\left(\frac{1}{2} - c_i\right)}{\ln\left(\frac{c_i}{1-c_i}\right)}. \quad \text{(E13)}$$

### 2. Diffusion

From irreversible thermodynamics follows that a gradient in chemical potential $\mu_H$ induces a current $\mathbf{J}$ of hydrogen such that

$$\mathbf{J} = -L\nabla\mu_H. \quad \text{(E14)}$$

At a given temperature, the chemical potential $\mu_H$ is only a function of the hydrogen concentration $c$ as shown in Eq. (E2). Thus the Eq. (E14) reduces to

$$\mathbf{J} = -D(c)\nabla c, \quad \text{(E15)}$$

with

$$D(c) = L\frac{d\mu_H}{dc}. \quad \text{(E16)}$$

We do not know *a priori* the function $L = L(c)$, however the lattice gas expression

$$\frac{d\mu_H}{dc} = \frac{RT}{c(1-c)} - A, \quad \text{(E17)}$$

from Eq. (E5) can be used to determine the behaviour of $L = L(c)$ in two limiting cases. From Fick's law [23], we know that the diffusion coefficient $D(c)$ in Eq. (E1) tends towards a constant for $c \to 0$ or $c \to 1$. In the limit $c \to 0$,

$$D_0 = D(c)\big|_{c \to 0} = L\frac{d\mu_H}{dc}\bigg|_{c \to 0} \cong L\frac{RT}{c}, \quad \text{(E18)}$$

hence

$$L_{c \to 0} = \frac{D(c)\big|_{c \to 0}}{RT}c = \frac{D_0}{RT}c. \quad \text{(E19)}$$

Similarly, in the limit $c \to 1$, we have

$$L_{c \to 1} = \frac{D(c)\big|_{c \to 1}}{RT}(1-c) = \frac{D_1}{RT}(1-c). \quad \text{(E20)}$$

Assuming for simplicity that

$$D_0 = D_1, \quad \text{(E21)}$$

we obtain from Eqs. (E14), (E19)–(E21)

$$\mathbf{J} = -D_0\left[1 - \frac{A}{RT}c(1-c)\right]\nabla c. \quad \text{(E22)}$$

This equation implies that the diffusion coefficient

$$D(c) = D_0 f(c) = D_0\left[1 - \frac{A}{RT}c(1-c)\right], \quad \text{(E23)}$$

is strongly dependent on the H concentration $c$. This is fully in agreement with the detailed measurements on Nb-H discussed extensively by Alefeld *et al.* [50] The function $f(c)$ in Eq. (E23) can also be expressed in terms of the critical temperature in Eq. (E9). Then

$$f(c) = \left[1 - 4\frac{T_c}{T}c(1-c)\right]. \quad \text{(E24)}$$

At the critical concentration $c_c = 1/2$, the diffusion coefficient tends towards zero when the temperature is decreased to the critical temperature $T_c$. At the critical temperature Eq. (E24) reduces to

$$f_c(c) = 1 - 4c(1-c). \quad \text{(E25)}$$

The concentration dependence predicted by Eq. (E24) for various values of the parameter $T/T_c$ are shown in Fig. 15.

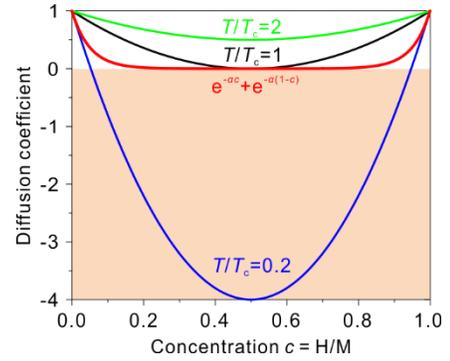

FIG. 15. Concentration dependence of the diffusion coefficient predicted by the lattice-gas expression in Eq. (E24) for three temperatures $T$ normalized to the critical temperature $T_c$. The red curve represents the concentration dependent diffusion coefficient function $[e^{-\alpha c} + e^{-\alpha(1-c)}]$ actually used for the simulations. At low $c$ and $c \approx 1$, the red curve has the same slope as the parabola (blue) with $T/T_c = 0.2$.

Although Eq. (E25) has been used to model Li diffusion in Li-ion batteries [51], it is not applicable to the simulation of the finger patterns in this work. This is easily understood since our experiments are carried out at temperatures where the dilute α-MH$_c$ and concentrated β-MH$_c$ coexist. This corresponds to a situation in our model with $T \ll T_c$ and thus Eq. (E23) implies negative diffusion coefficients for concentrations $c_s^{low} < c < c_s^{high}$, where $c_s^{low}$ and $c_s^{high}$ are the two solutions of Eq. (E11). We found that an efficient way to simulate diffusion fronts is to use the following concentration dependent diffusion coefficient

$$D(c) = D_0 f(c) = D_0\left[e^{-\alpha c} + e^{-\alpha(1-c)}\right]. \quad \text{(E26)}$$

Good agreement with experiment is obtained with $\alpha = 20$. The very small value of $D(c)$ for concentrations $\frac{1}{\alpha} < c < 1 - \frac{1}{\alpha}$ leads to a drastic slowing down of the diffusion process, and thus to a very steep front, in agreement with the experimental data in Figs. 3 to 5. The good agreement obtained with the observed finger patterns is obviously an indication of the usefulness of Eq. (E26). It is furthermore important to point out that the same concentration dependent diffusion coefficient explains also the observed $H_2$ pressure dependence of the front mobility described in Appendix D.

### 3. Simulation procedure

Our simulations are based on a diffusion equation. Particle conservation

$$\frac{\partial c}{\partial t} + \nabla \cdot \boldsymbol{J} = 0, \tag{E27}$$

together with Eqs. (E15), (E26), and (E27), lead to the non-linear diffusion partial differential equation

$$\frac{\partial c}{\partial t} = D_0 \nabla \cdot \left[ f(c) \nabla c \right]. \tag{E28}$$

In our numerical procedure, this equation is solved using Euler integration with for hydrogen loading a $c = 1$ boundary condition at the Pd entrance gate and zero flux boundary conditions elsewhere at the borders of the sample. For unloading, a $c = 0$ boundary condition is used. For the numerical algorithm, Eq. (E28) is rewritten as

$$\frac{\partial c}{\partial t} = D_0 \nabla f(c) \cdot \nabla c + D_0 f(c) \nabla^2 c. \tag{E29}$$

The first order numerical derivatives yield values in between the cells that tile the sample and hence are averaged between adjacent inter cell values, thus yielding values at the cells. This (together with a not too large time step) prevents the well-known unphysical generation of checkerboard patterns. The second order numerical derivative automatically gives values at the cells and is adjusted to prevent the generation of rectangular patterns that reflect the cells by a method analogous to that of Oono *et al*. [45]

### 4. Loading-unloading symmetry in diffusion simulations

For a deeper understanding of the H loading/unloading symmetry, we indicate in Fig. 16(a) how the concentration profiles evolve during H loading with a fixed concentration $c = 1$ applied at the left-hand side of the sample until the sample is half filled with hydrogen. Then the concentration on the left-hand side is abruptly decreased to 0 and the concentration profiles evolve as indicated in Fig. 16(b). Two features are remarkable: i) the front moves in both cases in the same way, *i.e.*, from left to right, although the concentration gradient at the H loading/unloading diffusion front is reversed; ii) the diffusion front at the end of H loading time (indicated by the dash line in Fig. 16) remains essentially unchanged until it is hit by the unloading front (from left-hand side). This is exactly what is observed experimentally in Fig. 12.

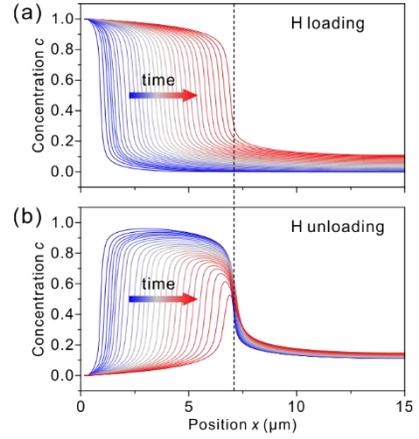

FIG. 16. Time evolution of calculated concentration profiles (a) during H loading with an imposed boundary concentration equal to 1 on the left-hand side, and (b) after a sudden decrease of the boundary concentration to 0. In both cases, there is no particle current at the right-hand side of the sample. The arrows indicate that the front movements are the same in both cases. The dash line indicates the diffusion front position at the end of H loading time.

### 5. Diffusion simulations for samples with inserted objects

We describe here in detail the procedure followed to generate the simulated fronts in Figs. 4(c) and 5(b). For example, we consider the situation where a prism (*i.e.*, Pd) is inserted in a medium (*i.e.*, Mg) as shown in Fig. 3(a). For the medium, we set $D_0$ in Eq. (E26) equal to unity, *i.e.*, $D_{\text{medium}} = 1$, while for the object we choose $D_{\text{object}} = 100$, so that $D_{\text{object}}/D_{\text{medium}} = 100$. In Fig. 17(a), a simulated concentration profile for this situation is shown at a time when the prism is partially filled with H. The line of constant concentration $c = 0.5$ (thick blue lines in Fig. 17) is taken as the diffusion front at that time. Front lines for a set of selected times are shown as red lines in Fig. 17(b).

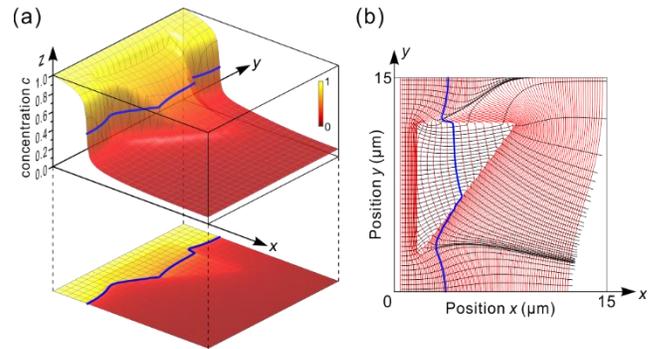

FIG. 17. (a) Concentration profile obtained from Eqs. (E26) and (E28) with $D_{\text{object}}/D_{\text{medium}} = 100$ at a time where the inserted prism with fast H diffusion is approximately half-filled with H. The blue line of constant concentration $c = 0.5$ is taken as the diffusion front line. (b) Front lines (red) for a large set of times $t_n$ that are equidistant in $t_n^{1/2}$. The corresponding streamlines (black) are locally perpendicular to the front lines.

## 6. Effective diffusion coefficient of a Pd object in quasi-free Mg

The simulations lead to streamlines that are consistent with the observed finger patterns measured at 353 K if the ratio

$$\frac{D_{object}}{D_{medium}} = 100 . \tag{E30}$$

Our object is made of Pd, while our medium is Mg. At 353 K the diffusion coefficient of H in Pd is [50]

$$D_{Pd}(353 \text{ K}) = 1.5 \times 10^{-10} \text{ m}^2\text{s}^{-1}, \tag{E31}$$

and from the measured value of $D_{Mg}$(353K), we expect naively

$$\frac{D_{Pd}}{D_{Mg}} = \frac{1.5 \times 10^{-10} \text{ m}^2\text{s}^{-1}}{3.1 \times 10^{-15} \text{ m}^2\text{s}^{-1}} \approx 4.8 \times 10^4 . \tag{E32}$$

That the actual ratio in Eq. (E30) is much lower is due to the fact that as soon as the MgH$_2$/Mg front hits the Pd object (*i.e.*, prism, concave or convex lenses), H atoms do not only diffuse in the Pd, but also leak constantly towards the medium above it. We have then a bi-layer situation as considered by de Man *et al.* [52] Using the same model for a Pd/Mg bilayer we obtain

$$K_{object} = 2D_{object} \frac{c'_{Pd}}{c'_{Mg}} \frac{d_{Pd}}{d_{Mg}}, \tag{E33}$$

where $c'_{Pd}$ is the concentration of H (unit: mol cm$^{-3}$) when the front is reaching the left side of the object. For the samples in Figs. 1 to 5, $d_{Pd}$ = 20 nm and $d_{Mg}$ = 45 nm. The concentration in the Mg is taken as

$$c'_{Mg} = \frac{2}{13.98} \text{ mol cm}^{-3} . \tag{E34}$$

The front mobility at the object is thus a function of the concentration $c'_{Pd}$

$$K_{object} = \frac{2D_{Pd} c'_{Pd}}{c'_{Mg}} \frac{20 \text{ nm}}{45 \text{ nm}} = \frac{9.3 \times 10^{-10} \text{ m}^2\text{s}^{-1}}{\text{mol cm}^{-3}} c'_{Pd} . \tag{E35}$$

From the observation that the experimental fronts are well reproduced with $D_{object}/D_{medium}$ = 100, we conclude that

$$K_{object} = 100 K_{medium} = 100 \times 3.1 \times 10^{-15} \text{ m}^2\text{s}^{-1}, \tag{E36}$$

and consequently that

$$c'_{Pd} = 3.3 \times 10^{-4} \text{ mol cm}^{-3}, \tag{E37}$$

which corresponds to (considering the molar volume of Pd, $V_{Pd} = 8.85 \text{ cm}^3\text{mol}^{-1}$)

$$c_{Pd} = \frac{\text{H}}{\text{Pd}} = c'_{Pd} V_{Pd} = 2.9 \times 10^{-3} . \tag{E38}$$

This is the H concentration at the entrance of the Pd object when the MgH$_2$/Mg front in the medium hits the Pd object. It is thus the H/Pd at a pressure equal to the plateau pressure of Mg at 353 K which is approximately 0.1 kPa. The low value for the concentration H/Pd in Eq. (E38) is fully consistent with the hydrogenography data of Pivak *et al.* [53] on free standing Pd films.

## 7. Model for the pressure dependence of front mobility

The purpose of this part is to demonstrate that the remarkable increase of the front mobility with increasing H$_2$ pressure shown in Fig. 14(b) can fully be understood by means of this model without introducing any new fitting parameter. It turns out to be a direct consequence of the concentration dependence of the diffusion coefficient in Eq. (E26). As the applied hydrogen pressure $p$ determines the H concentration $c_{boundary}$ at the entrance gate of the sample, one expects that the front mobility $K$ depends on $p$ via the concentration dependence of the diffusion coefficient.

In the first step we use a 1-dimensional simulation to determine the connection between $c_{boundary}$ and $K_{simulation}$. The simulations done with

$$D(c) = e^{-20c} + e^{-20(1-c)}, \tag{E39}$$

lead to the results in Fig. 18(a). There is indeed a strong variation of the simulated mobility $K_{simulation}$ with the boundary concentration, which is accurately described by

$$K_{simulation} = \exp(23.35 \, c_{boundary} - 25.72) . \tag{E40}$$

In the second step, we use Eq. (E5) to relate the applied H$_2$ pressure $p$ to $c_{boundary}$

$$\ln p = 2\ln\left(\frac{c_{boundary}}{1-c_{boundary}}\right) + 2\frac{\Delta H_\infty - Ac_{boundary} - T\Delta S}{RT} . \tag{E41}$$

As shown in Fig. 15, the red curve that corresponds to Eq. (E39) has the same slopes at low and high $c$ as the parabola with $T/T_c = 0.2$. From Eqs. (E9) and (E41) follows then that

$$\ln p = 2\ln\left(\frac{c_{boundary}}{1-c_{boundary}}\right) + 2\frac{\Delta H_\infty}{RT} - 40c_{boundary} - 2\frac{\Delta S}{R} . \tag{E42}$$

For high concentrations, when $c_{boundary}$ is close to 1,

$$\ln p \cong -2\ln(1-c_{boundary}) + 2\frac{\Delta H_\infty}{RT} - 40 - 2\frac{\Delta S}{R}$$
$$= -2\ln(1-c_{boundary}) + X , \tag{E43}$$

where

$$X = 2\frac{\Delta H_\infty}{RT} - 40 - 2\frac{\Delta S}{R}, \tag{E44}$$

is constant as all experiments are carried out at a constant temperature of 353 K. Solving Eq. (E43) for $c_{boundary}$ and introducing it into Eq. (E40) gives

$$K_{simulation} = \exp\left[23.35\left(1 - \frac{1}{\sqrt{p}} e^{\frac{1}{2}X}\right) - 25.72\right] . \tag{E45}$$

As the measured front mobility $K$ is equal to $K_{simulation}$ up to a multiplicative factor $K_0$, we expect for our experimental data

$$K = K_0 \exp\left[23.35\left(1 - \frac{1}{\sqrt{p}} e^{\frac{1}{2}X}\right) - 25.72\right], \tag{E46}$$

and thus that a plot of $\ln K$ versus $1/\sqrt{p}$ should be a straight line

$$\ln K = \frac{a}{\sqrt{p}} + b, \tag{E47}$$

with

$$a = -23.35 \exp\left(\frac{1}{2} X\right), \tag{E48}$$

and

$$b = -2.37 + \ln K_0. \tag{E49}$$

The experimental data points in Fig. 14(b) obey nicely the linear relation predicted by Eq. (E47). The slope $a = -0.07$ and the intercept $b = -33.3$ lead to

$$K = \exp\left(\frac{-0.07}{\sqrt{p}} - 33.3\right), \tag{E50}$$

which corresponds to the blue curve in Fig. 18(b).

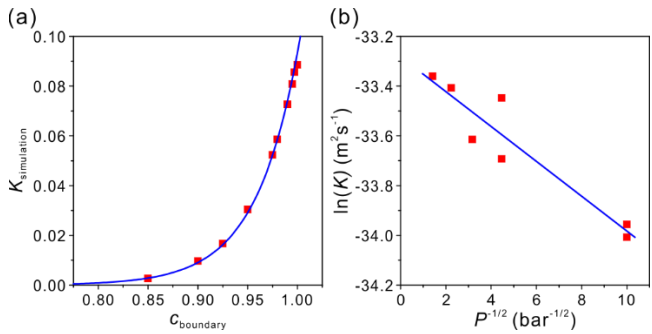

FIG. 18. (a) Influence of the boundary concentration on the front mobility. The blue curve is the fit to the calculated points given in Eq. (E40). (b) The same experimental data in Fig. 14(b) for the pressure dependence of the front mobility plotted according to Eq. (E47), which confirms the predicted linear dependence between $\ln K$ and $P^{-1/2}$.

**8. Implications for the pressure-composition isotherms**

From Eq. (E48) and the value for the parameter $a = -0.07$ we find

$$X = -11.63. \tag{E51}$$

From the definition of $X$ in Eq. (E44) and the values $\Delta H_\infty = 21 \text{ kJ} \left(\text{moleH}\right)^{-1}$ and $\Delta S = -67.5 \text{ JK}^{-1} \left(\text{moleH}\right)^{-1}$ from Ref. [19], we obtain

$$X = 2\frac{\Delta H_\infty}{RT} - 40 - 2\frac{\Delta S}{R} = -9.45 \tag{E52}$$

This is in remarkably good agreement with the experimental value in Eq. (E51) considering the simplicity of the model and the uncertainties in the thermodynamic values for the Mg-H system in the dilute regime.

*Permeability of Separation Membrane Materials*, J. Membr. Sci. **444**, 70–76 (2013).

[53] Y. Pivak, R. Gremaud, K. Gross, M. Gonzalez-Silveira, A. Walton, D. Book, H. Schreuders, B. Dam, and R. Griessen, *Effect of the Substrate on the Thermodynamic Properties of PdH$_x$ Films Studied by Hydrogenography*, Scr. Mater. **60**, 348–351 (2009).